\documentclass{iau} 
\usepackage{graphicx}

\title[IAU-S334~~NGC 5053: Orbit, Mg, Al, and Si] 
{On The Metal-poor non-Sagittarius Globular Cluster NGC 5053: Orbit, Mg, Al, and Si}

\author[Baitian Tang]   
{Baitian Tang$^{1}$}

\affiliation{$^{1}$Departamento de Astronom\'ia, Universidad de Concepci\'on, Casilla 160-C, Concepci\'on, Chile 
	 \\ email: {\tt tangbaitian@gmail.com}}

\pubyear{2017}
\volume{334}  
\jname{Rediscovering our Galaxy}
\editors{C. Chiappini, I. Minchev, E. Starkenburg, M. Valentini., eds.}
\begin{document}

\maketitle

\begin{abstract}
Metal-poor globular clusters (GCs) show intriguing Al-Mg and Si-Al correlations, which are important clues to decipher the multiple population phenomenon. NGC 5053 is one of the most metal-poor GCs, and has been suggested to be associated with the Sagittarius dwarf galaxy (Sgr), due to its similar location and radial velocity with one of the Sgr arms. In this work, we simulate the orbit of NGC 5053, and argue against the connection between Sgr and NGC 5053. Meanwhile, Mg, Al, and Si spectral lines, which are difficult to detect in the optical spectra, have been detected in the near-infrared APOGEE spectra. We use three different sets of stellar parameters and  codes to derive the Mg, Al, and Si abundances, and we always see a large Al variation, and a substantial Si enhancement. Comparing with other metal-poor GCs, we suggest metallicity may not be the only parameter that controls the multiple populations.

\end{abstract}

The multiple population (MP) phenomenon has been found in globular clusters (GCs) for almost half a century, but the astronomical community is still looking for a successful theory to explain all the observational evidences. Among them, the Al-Mg and the less frequent Si-Al correlations are key factors in decipher the MP phenomenon, because these correlations indicate high core temperature of the polluting stars (\cite[Ventura et al. 2003]{Ventura2013}), which is a strong constraint on the nature of the polluting stars. These correlations are more prominent in metal-poor GCs, therefore further investigation of metal-poor GCs will tell us more about the environment that stimulates the MPs. 

\begin{figure}
	\begin{center}
		\includegraphics[width=0.70\textwidth]{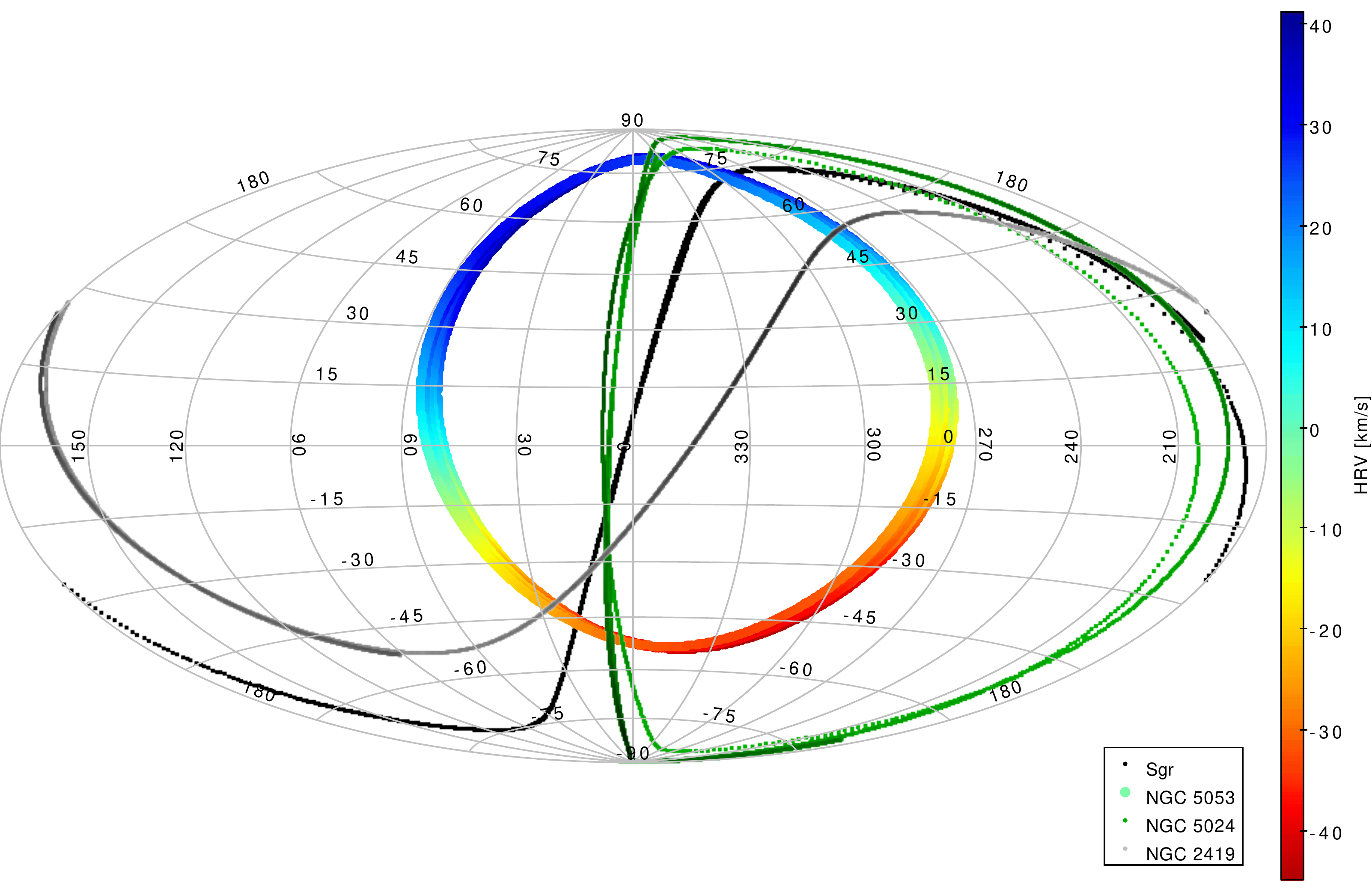}
		\caption{Simulated orbits of Sgr (black dots), NGC 5053, M53 (NGC 5024, green dots), and NGC 2419 (grey dots) using GravPot16. The orbit of NGC 5053 is color-coded by heliocentric radial veolocity. The orbits of NGC 5053 and M53 do not resemble that of Sgr, suggesting no association with Sgr.}		\label{Figure1}
	\end{center}
\end{figure}

NGC 5053 is one of the most metal-poor GCs ([Fe/H$]=-2.27$). With similar location and radial velocity (RV) to one of the Sagittarius (Sgr) arms (\cite[Law \& Majewski 2010]{Law2010}), NGC 5053 has been speculated to be associated with Sgr. In this work, we have simulated the orbit of NGC 5053 using our new dynamical software, GravPot16 (Fernandez-Trincado et al., in prep), where we assume the observed RV, distance, and HST proper motion (Tony Sohn, priv. com.) of NGC 5053. Our results show that the orbit of NGC 5053 does not resemble that of Sgr, arguing against a connection between them (Fig. \ref{Figure1}). 

The APOGEE survey employs the multi-object NIR fiber spectrograph on the 2.5 m telescope at Apache Point Observatory to deliver high-resolution ($R\sim$22,500) $H$-band spectra ($\lambda = 1.51 - 1.69$ $\mu$m). For the first time, strong Mg, Al, and Si spectral lines have been detected in the APOGEE spectra of the NGC 5053 cluster members. We have identified 10 stars with similar location, iron abundances and radial velocities, and these stars are further confirmed by their location in the color-magnitude diagram. Because the spectral lines are generally weak for metal-poor stars, manual analysis is suggested for abundance studies. Due to the uncertainties associated with the stellar parameters, we have adopted three different sets of stellar parameters (SPs) and two different codes to measure chemical abundances.

\begin{figure}
	\begin{center}
		\includegraphics[width=0.85\textwidth]{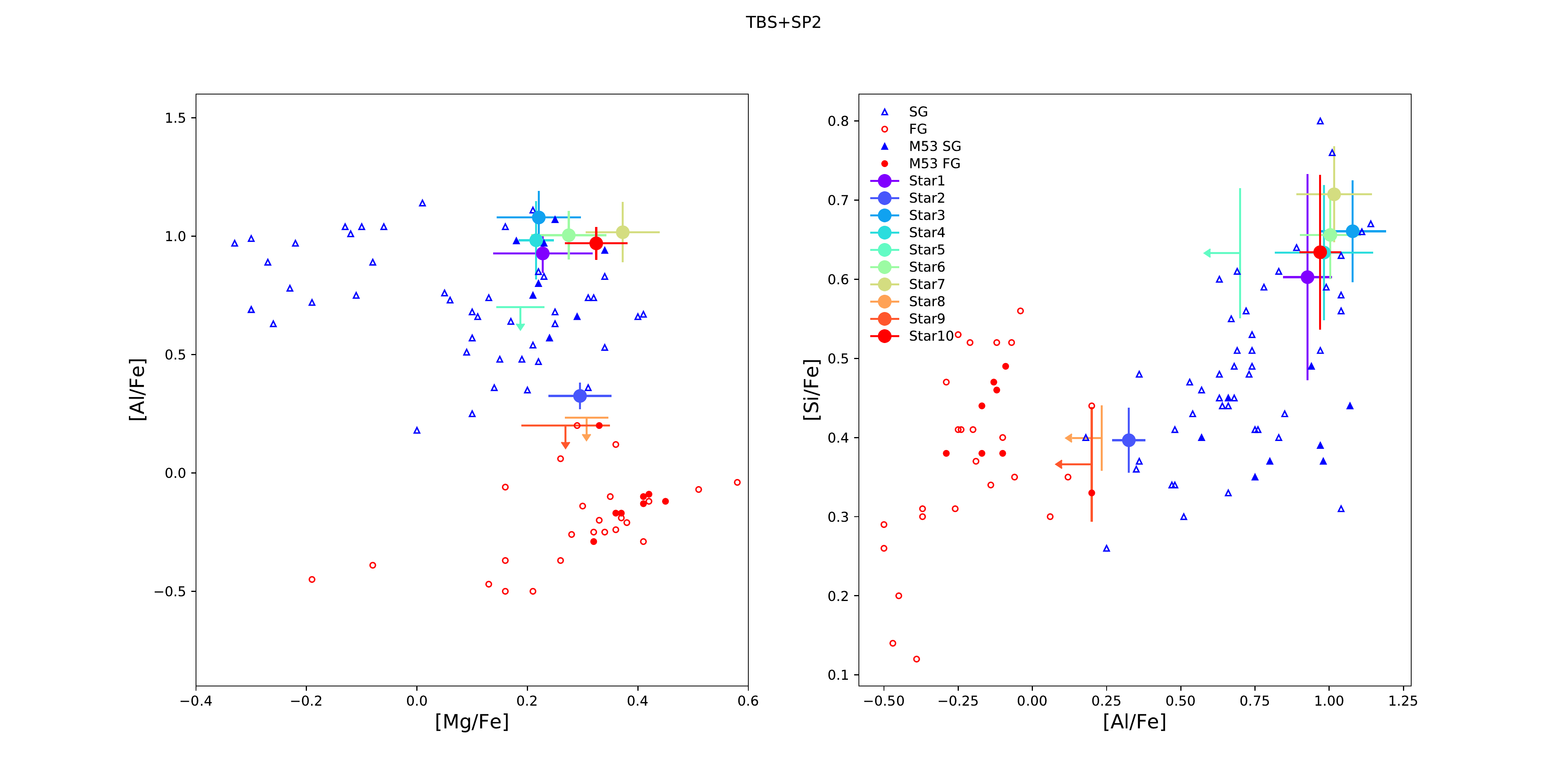}
		\caption{$Left$: [Al/Fe] versus [Mg/Fe]. $Right$: [Si/Fe] versue [Al/Fe]. The NGC 5053 stars are labelled as colored dots. The two generation stars in \cite[Meszaros et al. (2015)]{Meszaros2015} are labelled as blue and red triangles. Three stars with estimated Al upper limits are indicated by arrows.}		\label{Figure2}
	\end{center}
\end{figure}

Due to the page limit, we only show one set of SP and code combination in Fig. \ref{Figure2}. We will show more results in the coming journal paper. Basically, the general trends are the same for different methods, with possible detailed differences in the final chemical abundances. We see a large Al variation, and a substantial Si enhancement. Comparing with other metal-poor GCs, we suggest that metallicity may not be the only parameter that controls the multiple populations.

\textit{{\bf Acknowledgements:} B.T. gratefully acknowledges financial support from the IAU, and the Chilean BASAL Centro de Excelencia en Astrof\'isica y Tecnolog\'ias Afines (CATA) grant PFB-06/2007. This research is based on the SDSS-III/APOGEE survey spectra.}


\end{document}